\documentclass[aps,prl,preprint,groupedaddress,showpacs]{revtex4-1}
\usepackage{amsmath, amssymb,graphicx}
\usepackage{color}

\newcommand{\R}{\mathbb{R}}

\newcommand{\bigoh}{\mathcal{O}}

\newcommand{\rrj}{{ r_{i,j}^{\tau}}}
\newcommand{\rrjd}{{\dot{r}_{i,j}^{\tau}}}

\usepackage{color}
\begin{document}

%\preprint{}

\title{The impact of anticipation in dynamical systems}


\author{P. Gerlee$^1$, K. Tunstr\o m$^2$, T. Lundh$^1$ and B. Wennberg$^1$}
%\thanks{}
\affiliation{$^1$Mathematical Sciences, Chalmers University of Technology and University of Gothenburg, 412 96 G\"oteborg}
\affiliation{$^2$Department of Physics, Chalmers University of Technology, 412 96 G\"oteborg}


\date{\today}

\begin{abstract}
Collective motion in biology is often modelled as a dynamical system, in which individuals are represented as particles whose interactions are determined by the current state of the system. Many animals, however, including humans, have predictive capabilities, and presumably base their behavioural decisions---at least partially---upon an anticipated state of their environment. We explore a minimal version of this idea in the context of particles that interact according to a pairwise potential. Anticipation enters the picture by calculating the interparticle forces from linear extrapolations of the particle positions some time $\tau$ into the future. Simulations show that for intermediate values of $\tau$, compared to a transient time scale defined by the potential and the initial conditions, the particles form rotating clusters in which the particles are arranged in a hexagonal pattern. Analysis of the system shows that anticipation induces energy dissipation and we show that the kinetic energy asymptotically decays as $1/t$. Furthermore, we show that the angular momentum is not necessarily conserved for $\tau >0$, and that asymmetries in the initial condition therefore can cause rotational movement. These results suggest that anticipation could play an important role in collective behaviour, since it induces pattern formation and stabilises the dynamics of the system. 
\end{abstract}

%\keywords{}

\maketitle
%Introduction to collective behaviour
\section{Introduction}
Countless examples of collective motion are found in biological systems, spanning from swarming bacteria to human crowds~\cite{Sumpter2006}. 
The dynamical patterns exhibited by groups of animals have fascinated humans across millennia, but with modern technology, this fascination has been channeled into an active research area, where methodologies across disciplines---from biology to physics and engineering sciences---are essential to achieve progress \cite{Giardina_2008,Vicsek:2012ty,Lopez:2012vf}. 

%Interacting particle models of collective behaviour
From a modeling perspective, much research derives from the idea that simple rules of interaction between animals, e.g. attraction, repulsion and alignment~\cite{Couzin02}, can explain observed swarming patterns~\cite{Tunstrom2013}. Typically, swarming behaviour is modelled as a collection of particles that represent the organisms in question. To each particle one assigns a position and velocity. The velocity determines the evolution of the position, and the velocity is in turn influenced by the position and velocity of neighbouring particles. 
% Using such models it has been shown that distinct dynamical patterns can be caused by e.g.\ velocity alignment \cite{Cucker2007} and attraction between particles \cite{Strombom2011}. 
Taking inspiration from physics, the interactions between individuals are often assumed to result from pairwise forces that only depend on the distance between the individuals. Typically the force is defined via a potential that contains a repulsive and attractive region such that at some intermediate distance the interaction energy is minimised. In terms of animals this would correspond to some preferred distance between neighbouring individuals~\cite{Katz_2011}. In addition, most physics inspired models rely on self-propulsion to drive pattern formation~\cite{Dorsogna2006} and often include a noise term as well \cite{vicsek1995}.%, as models without this ingredient display no clear dynamical patterns.  

%The idea of prediction/anticipation in animals
{An underlying assumption in most current collective motion models is that individuals react and update their velocity according to the current state of other individuals. Contrary to this, we know that many animals including humans have the ability to anticipate movement, and act on predicted states. Humans anticipate the movement of visual cues, such as other people in a moving crowd, by extrapolating their positions in time \cite{Nijhawan1994}. This process occurs within the retina itself, and is in fact necessary if we are to respond to rapid visual cues, since the delay induced by phototransduction is on the order of 100 ms \cite{Berry1999}. In addition to performing extrapolation the retina also detects when its prediction fails and signals this downstream to the visual cortex \cite{Schwartz2007}. Anticipation is not restricted to humans, but has also been detected among insects \cite{Collett1978}, amphibians \cite{Borghuis2015} and fish \cite{Rossel2002}. Therefore anticipation most likely plays an important role in the behaviour of animals that engage in flocking and swarming.}

{A recent study of human interactions in crowds quantified the effect of anticipation by showing that the strength of physical interaction does not depend on distance, but on the time to collision \cite{Karamouzas2014}. This suggests that humans act in an anticipatory fashion and use current positions and velocities to extrapolate possible future collisions and update their current velocity to avoid such collisions. In terms of the above mentioned interaction potential one might then assume that individuals interact according to anticipated future positions and adjust their current velocities according to the predicted state of the system. This idea has been investigated by Morin et al.\ \cite{Morin2015}, who considered the continuous-time Viscek model, that assumes self-propelled particles that interact with neighbouring particles within a certain interaction radius. The classical model was modified so that the difference in angle between particle $i$ and $j$, $\theta_i - \theta_j$,  was altered to $\theta_i - (\theta_j + \alpha \sigma_j)$, where $\sigma_j$ is the sign of the angular velocity of particle $j$ and $\alpha$ is some positive parameter. Depending on the value of $\alpha$ and the magnitude of the noise that model can exhibit isotropic behaviour, spinning and flocking, which shows that anticipation indeed can have important consequences. The effect of anticipation has also been investigated in a lattice-based model inspired by the swarming of soldier crabs \cite{murakami2017}. They showed that mutual anticipation leads to dense collective motion with a high degree of polarisation, and that the turning response depends on the distance between two individuals rather than the relative heading, which is in agreement with empirical data.}

%Describe the idea informally
%Introduce the model
In this paper we also explore the idea of anticipation, but in the context of an interacting particle system where the interaction forces are calculated not from current positions, but from positions extrapolated some time $\tau$ into the future. For the sake of simplicity we assume that the future positions are given by a linear extrapolation from the current velocities, although more elaborate means of extrapolation are conceivable (see Discussion). {In order to get a thorough understanding of the dynamics that anticipation induces we disregard self-propulsion, alignment and other processes commonly found in models of collective behaviour and focus on the effects in a simple of model where individuals are represented as  interacting particles.}

\section{A model of anticipation in dynamical systems}
To introduce the concept of anticipation let us begin with a simple example. Consider two particles of mass $m$ connected by a spring with rest length zero and spring constant $k/2$, where $k>0$. The distance $x$ between the two particles obeys the equation $m\ddot{x} = -kx$, which is simply that of a harmonic oscillator. Now let us assume that the particles anticipate the movement of one another, and hence that the forces acting on the particles are not given by their instantaneous positions, but by the anticipated positions some time $\tau$ in the future, i.e. we calculate the forces based on predicted positions ${x_i^p}(t+\tau) = x_i(t)+\tau v_i(t)$ for particle $i=1$ and 2. In this case the equation of motion of the interparticle distance $x$ is given by $m \ddot{x} = -kx -k\tau \dot{x}$. This is the equation for a damped harmonic oscillator, and hence we conclude that in this simple system anticipation has a damping effect on the dynamics. In contrast to the undamped system there is dissipation of energy and for any initial condition the system will reach a stationary state with $x=0$ as $t \rightarrow \infty$.

We now extend the idea of anticipation to $N$ identical particles that interact via some potential $U(r)$ and hence obey the following equations of motion: 
\begin{align}\label{eq:x}
\dot{\vec{x}}_i(t) &= \vec{v}_i(t), \\
%m\dot{\vec{v}}_i(t) &= -\vec{\nabla}_i U_i ({\vec{x}^p}(t+\tau)) \nonumber
m\dot{\vec{v}}_i(t) &= -\sum_{j\ne i} \vec{\nabla} U\left({\vec{x}_i^p}(t+\tau)-{\vec{x}_j^p}(t+\tau)\right), \nonumber
\end{align}
where $m=1$ is the mass and $\vec{x}_i, \vec{v}_i \in \R^2$, i.e. the system is defined in the plane. Again we assume that the individuals make use of a linear prediction of future positions such that $\vec{x}_i^p(t+\tau)=\vec{x}_i(t) + \tau\vec{v}_i(t)$. We note that for $\tau=0$ the system reduces to a standard system of interacting particles. 

{\color{black}
%Here are some comments to this:

%First, we cannot write $x_i(t+\tau) = x_i(t) + \tau v_i(t)$, because this is only correct if $x_i$ has constant velocity. We should use another symbol for the predicted value of $x_i$. I suggeset that we say something like:

%We calculate the forces based on a predicted future position, $\tilde{x}_i(t+\tau)$, where the prediction is calculated assuming constant velocity, $\tilde{x}_i(t+\tau)= x_i(t) + \tau v_i(t)$.

%The same kind of notation should be used also in the remaining part to avoid that other readers make the same mistake as I did in my previous reading. {\em If} we really use $x_i(t+\tau)$, then we have a functional differential equation that in the linear case could be treated by analysing the dispersion relation.

%My next comment is that this is really a dissipative system, as the following little calculation shows: 

}

%The interaction potential $U$ defines the forces acting between pairs of  particles, and we will use a generalised More %potential

{\color{black}
We will use a generalized Morse potential,
%\begin{align}
%&U_i(\vec{x}^p(t+\tau))= \\
%&\sum_{j=1,j\neq i}^N \left( C_r e^{-\vert \vec{x}^p_i(t+\tau)-\vec{x}^p_j(t+\tau)\vert/l_r } - C_a e^{-\vert \vec{x}^p_i(t+\tau)-\vec{x}^p_j(t+\tau)\vert/l_a }\right)
%\end{align}
\begin{align}
&U(x)= C_r e^{-\vert x\vert/l_r } - C_a e^{-\vert x\vert/l_a }\,,
\end{align}
to define the interaction between pairs of particles, where $C_r$ and $C_a$ represent the amplitude of the repulsive and attractive component of the potential, and $l_r$ and $l_a$ denote their respective ranges. In the simulations presented here we } will use $C_r=15$, $C_a=2.5$ and $l_a = 0.1$, $l_r=0.05$. We note that the force between particle $i$ and $j$ is in the direction of the vector %, which according to the classification of \cite{Dorsogna2006} puts us in the H-stable regime (region IV). 
\begin{align}
%\vec{r}_{ij} = \frac{\vec{x}_{i}(t)-\vec{x}_{j}(t) + \tau(\vec{v}_{i}(t)-\vec{v}_{j}(t))}{\vert \vec{x}_{i}(t)-\vec{x}_{j}(t) + \tau(\vec{v}_{i}(t)-\vec{v}_{j}(t)) \vert}.
r_{i,j}^\tau = \vec{x}_{i}(t)-\vec{x}_{j}(t) + \tau(\vec{v}_{i}(t)-\vec{v}_{j}(t)).
%\rrj=\vec{x}_i-\vec{x}_j+\tau(\vec{v}_i-\vec{v}_j)$
\label{eq:rij}
\end{align}
This implies that for $\tau > 0$ the direction of the force is generally not parallel to the vector $\vec{x}_i(t)-\vec{x}_j(t)$, pointing from particle $i$ to $j$, but is influenced by the velocity of the two particles.% (see fig.\ \ref{fig:schema}). 

In the following we will consider particles moving in two dimensions with open boundaries. We typically initialise the system with the particles at random positions within a disk, and zero initial velocities. All numerical solutions have been carried out using a Runge-Kutta method of order 8 from the GNU Scientific Library \cite{GNU}. 

% and integrate it numerically using the velocity verlet method with a time step of $\Delta t =10^{-3} $ \cite{Swope1982} with an open boundary.

\section{Results}
Before investigating the impact of anticipation on the dynamics we need to set the anticipation time $\tau$ into relation with other time and length scales present in the system. The model has one natural spatial scale, given by the minimum of the
Morse potential,
\begin{align}
  \bar{r}_p &= \mathrm{argmin}\ U(r)\,,
\end{align}
which with the Morse potential and parameters used here is given by
\begin{align}
 \bar{r}_p=\frac{l_a l_r}{l_a-l_r}\log\frac{C_r/C_a}{l_r/l_a} \approx 0.25.
\end{align}
The anticipation time, $\tau$, gives one natural time scale, but 
because the system is defined in the plane without boundaries
it is difficult to define time scales or spatial scales valid
asymptotically for long time intervals, to compare $\bar{r}_p$ and $\tau$
with. However, it is possible to define a relevant transient time scale
expressed in terms of the initial data. We let
%\begin{align}
%  \bar{r}_T&= {\mathbb E}[|\vec{x}_j-\vec{x}_k|]\,,
%\end{align}
%, and in the same spirit we
%let
\begin{align}
  \bar{f}_T&= {\mathbb E}[|\nabla U(|\vec{x}_j-\vec{x}_k|)|]\,,
\end{align}
be the expected interparticle force, where the expectation is taken over a randomly chosen pair of initial positions of the particles. From this we define a transient time scale
\begin{align}\label{eq:tt}
  \bar{t}_{T}&= \sqrt{2 \bar{r}_p / \bar{f}_T}\,,
\end{align}
which is the time needed for a particle with initial velocity zero to
move a distance $\bar{r}_p$ when subject to a constant force of strength $\bar{f}_T$. %A different definition would be to compute the net force on each particle, and to compute the average of their magnitudes, but at least with random initial data, the two definitions give similar results. 
%We have carried out simulations with various ratios of $\bar{r}_T/\bar{r}_p$ and
%$\tau/\bar{t}_{T}$, and present some of these results below.

By solving the equations of motion \eqref{eq:x} numerically with initial positions uniform random within a disk and initial velocities equal to zero, we observe for $\tau/\bar{t}_{T} \approx 1$, independent of particle number, the rapid formation of milling structures, where the particles organise into rotating patterns. For smaller $\tau/\bar{t}_{T} \ll 0.1$ the particles behave much like the classical system where nearby particles tend to form clusters, whereas others disperse. For $\tau/\bar{t}_{T} \gg 10$ particles also disperse, but do so in smaller clusters of aligned particles. The following section are devoted to understanding this behaviour for both small and large particle numbers. 

%We return to these properties in section X, but first we focus on the dynamics of small systems in the regime of prediction times where collective behaviour emerges. 
%this also occurs with the difference that the particles disperse not as single individuals, but in the shape of coherent groups 
%(see fig.\ \ref{fig:tau}). 
\subsection{Small particle numbers}
To get a better understanding of the effect of anticipation we start by looking at the dynamics of small systems with $N \in [2,11]$ and set  $\tau = 1$. The particles are placed at random positions in the unit circle and the initial velocities are zero for all particles. This implies that the initial density and therefore the transient time scale will vary with $N$, but for small $N$ we have that $\tau$ and $\bar{t}_{T}$ are of same order of magnitude, and hence we expect collective behaviour to emerge.

For $N=2$ the system behaves much like the two-particle system mentioned in the introduction. The particles attract one another and rapidly converge to a distance given by the equilibrium distance of the pairwise potential. This suggests that energy is being dissipated, a fact we will return to later. 

\begin{figure}[!htb]
\centering
\includegraphics[width=\columnwidth]{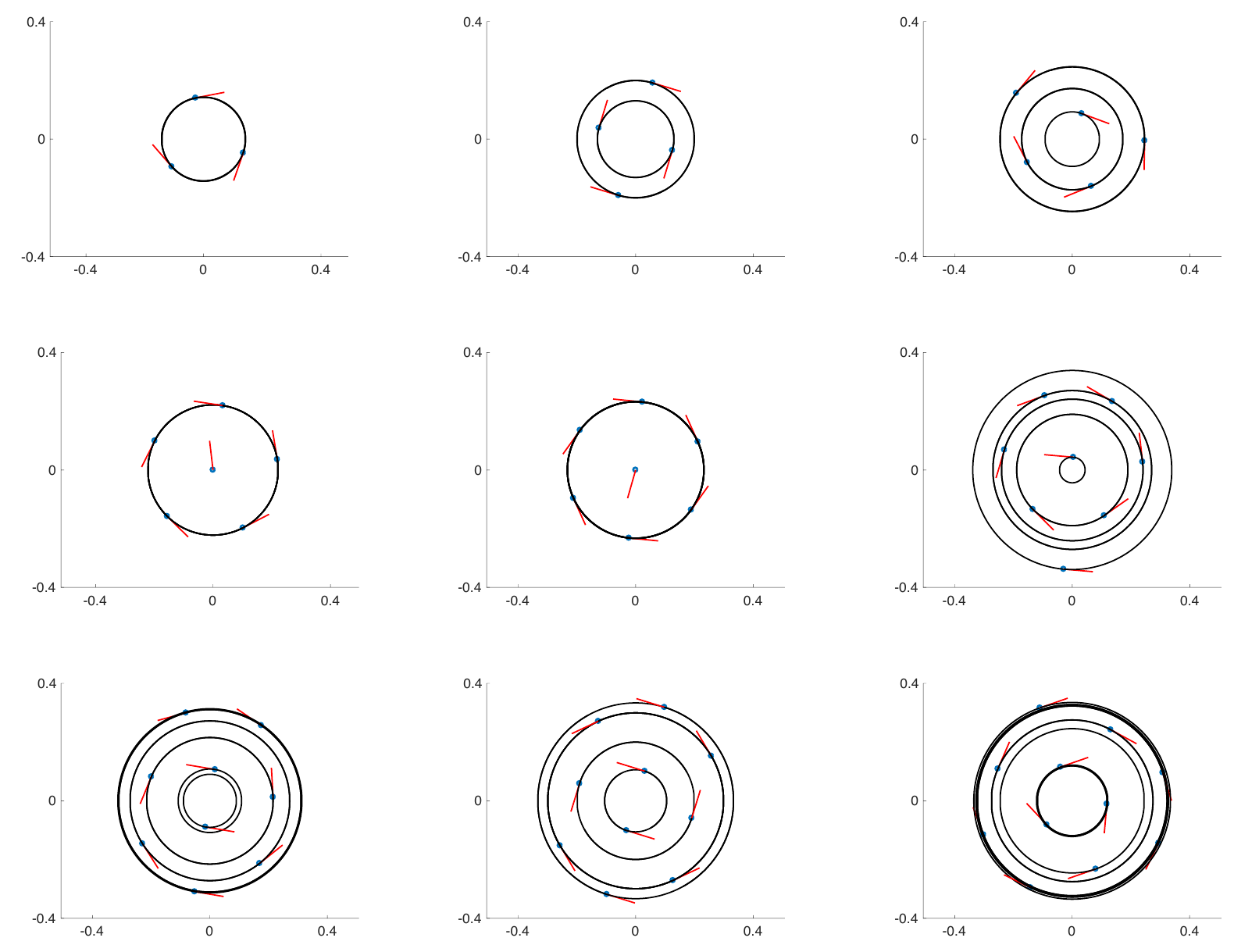}
\caption{\label{fig:smallN}{
%\obs{Larger axes ticks fonts both main plot and inset. Let y-max be about 1.05 to allow full graph to be shown.}
Rotational configurations observed for $N \in [3,11]$. Particles are shown as circles and red lines correspond to the unit velocity vector of each particle. The lines show the trajectories of the particles where the first 10 time steps have been discarded.}}
\end{figure}

For $N=3$, the particles start rotating on the same orbit around their common centre of gravity. A similar behaviour is seen for larger systems, but the number of unique orbits and their radii depend on the number of particles $N$ (see fig.\ \ref{fig:smallN}).  We note that the configurations closely resemble a partial hexagonal lattice built from equilateral triangles. These patterns are rigid rotations of a locally crystalline configuration, much like the dynamics reported in region VI of \cite{Dorsogna2006}. The main difference being that in our case there is no self-propulsion, and the particles are only influenced by a pairwise potential with anticipation. 

%For larger systems the particles also form rotating structures, in which the particle configuration depends on the particle number $N$ (see fig.\ \ref{fig:smallN}). The particles move along circular orbits around the centre of gravity of the structure.% and we have numerically confirmed that these configurations minimise the total potential energy of the system.

\subsubsection{Analysis of milling patterns}
\begin{figure}[!htb]
   \centerline{\includegraphics[width=\columnwidth]{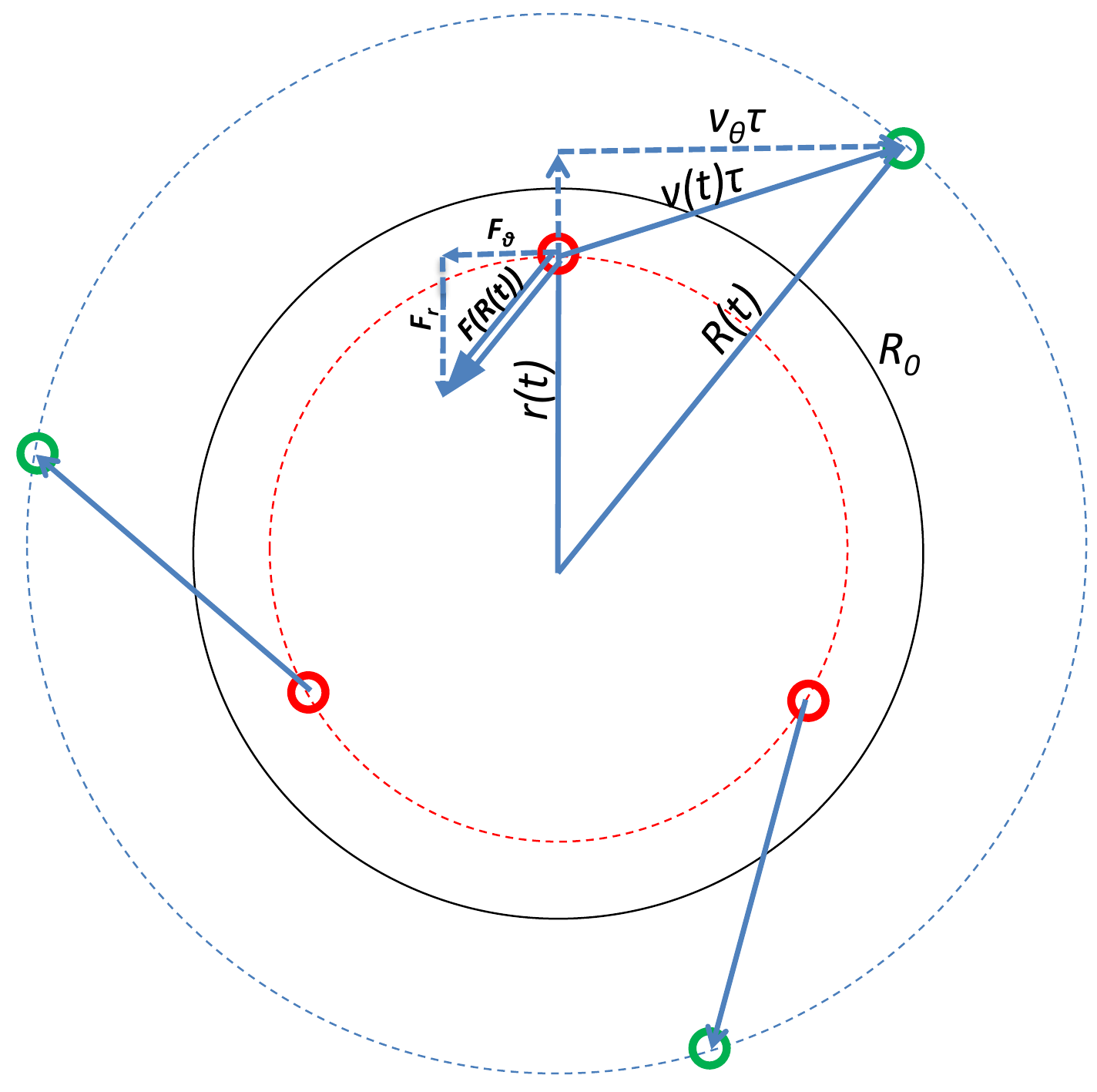}}
\caption{\label{fig:schema}
%\obs{Decrease font size somewhat.}
A schematic picture of a symmetric three particle system that is rotating on a circle with a radius $r(t)$ which is smaller than the equilibrium radius $R_0$. The small red circles show the current positions and the green circles denote the anticipated positions obtained by adding the vector $\vec{v}(t) \tau$ to the current position, i.e. $\vec{R}(t) = \vec{r(t)}+\vec{v}(t) \tau$. Since the anticipated position is located outside the equilibrium circle, there will be an attracting force applied at the actual position $\vec{r}(t)$. By decomposing this force, we get a central acceleration and a tangential retardation that eventually makes the speed of the particles go to zero.}% according to this asymptotic function $v_\theta(t)= \frac{1}{\sqrt{\frac{2 \tau t}{R_0^2} + \frac{1}{v_\theta(0)^2}}}$ .}
\end{figure}

The reason behind the observed rotation for $N>2$ can be understood by considering the forces acting on each particle (see fig.\ \ref{fig:schema}). For simplicity we consider the case $N=3$. We assume that the three particles are symmetrically distributed, i.e.\ with angle $\frac{2 \pi}{3}$ between them, on the inner dashed circle of radius $r(t)$ and with the same speed $|\vec{v}(t)|$ in the clockwise direction. For generality we assume that the particles move with velocity $\vec{v}(t)$ which has both a tangential and radial component. The non-zero velocity gives rise to anticipated positions that lie on a circle with radius $R(t)$. A third radius is also of importance, $R_0$, which is the radius at which the interparticle distance equals the equilibrium distance of the potential. For the case $N=3$, we have that
the equilibrium (or resting) radius is
\[R_0=\frac{l_a l_r \log(\frac{C_r l_a}{C_a l_r})}{\sqrt{3} (l_a - l_r)}.\]

A symmetric configuration on $R_0$ implies that the force from the potential $\vec{F}(R_0)=0$, where $\vec{F}(R)$ is the net force acting on each particle when at a distance $R$ from the centre. If $R(t) > R_0$ the net force will be radially inwardly directed from the point $\vec{R}(t)$. Now, this force, $\vec{F}(R(t))$ is not applied on the point $\vec{R}(t)$ which is the anticipated position on the outer circle, but instead at the current positions at $\vec{r}(t)$. By decomposing this force into its radial and tangential components we get a triangle which is similar to the right-angled triangle which has $\vec{R}(t)$ as the hypotenuse (see fig.\ \ref{fig:schema}). The net force can be decomposed into a radial component, $F_r$, that is generating a central force, and a tangential force $F_\theta$ which is retarding the rotation. Due to the fact that the anticipated position is outside $R_0$, the particles will experience attractive forces, although they in fact are located in the repulsive region of the potential. However, note that this is only true if $|\vec{v}(t)|\tau$ is large enough so that the anticipated position lies outside of $R_0$.

%%%

To obtain a quantitative understanding of the three particle system we formulated an ODE-system for the special case of $N$ particles located on a circle (as in figure \ref{fig:schema}). We describe the dynamics of a single particle, but since the configuration is symmetric the positions and velocities of the other particles can be obtained by shifting the solution appropriately. The equations of motion are given by:
 
\begin{equation}\label{eq:odes}\left\{
\begin{array}{ll} \dot{\vec{r}}(t) = \vec{v}(t) \\
\dot{\vec{v}}(t) = - F(R(t)) \frac{\vec{R}(t)}{R(t)} \\
\end{array}\right.
\end{equation} 
where $\vec{R}(t)=\vec{r}(t)+\tau \vec{v}(t)$ and $R(t) = |\vec{R}(t)|$. The force is given by
\begin{equation}
F(R(t))=
2 \sum_{k=1}^{\frac{n-1}{2}} \biggl( \frac{C_a}{l_a} e^{-\alpha_k/l_a} - \frac{C_r}{l_r} e^{-\alpha_k/l_r}\biggl) \cos(\frac{\pi-k \frac{2\pi}{n}}{2})%, \;\; \mbox{ if $n$ is odd.}
\end{equation}
where
\begin{equation}
\alpha_k= \sqrt{2}R(t)\sqrt{1-\cos(k \frac{2\pi}{n})}
\end{equation}
is the interparticle distance between the focal particle and particle $k$.
%For the special case where $n=3$, we have that the equilibrium (resting) radius is
%\[R_0=\frac{l_\alpha l_r \log(\frac{C_r l_\alpha}{C_\alpha l_r})}{\sqrt{3} (l_\alpha - l_r)}.\]
 
\begin{figure}[!htb]
\centering
\includegraphics[width=\columnwidth]{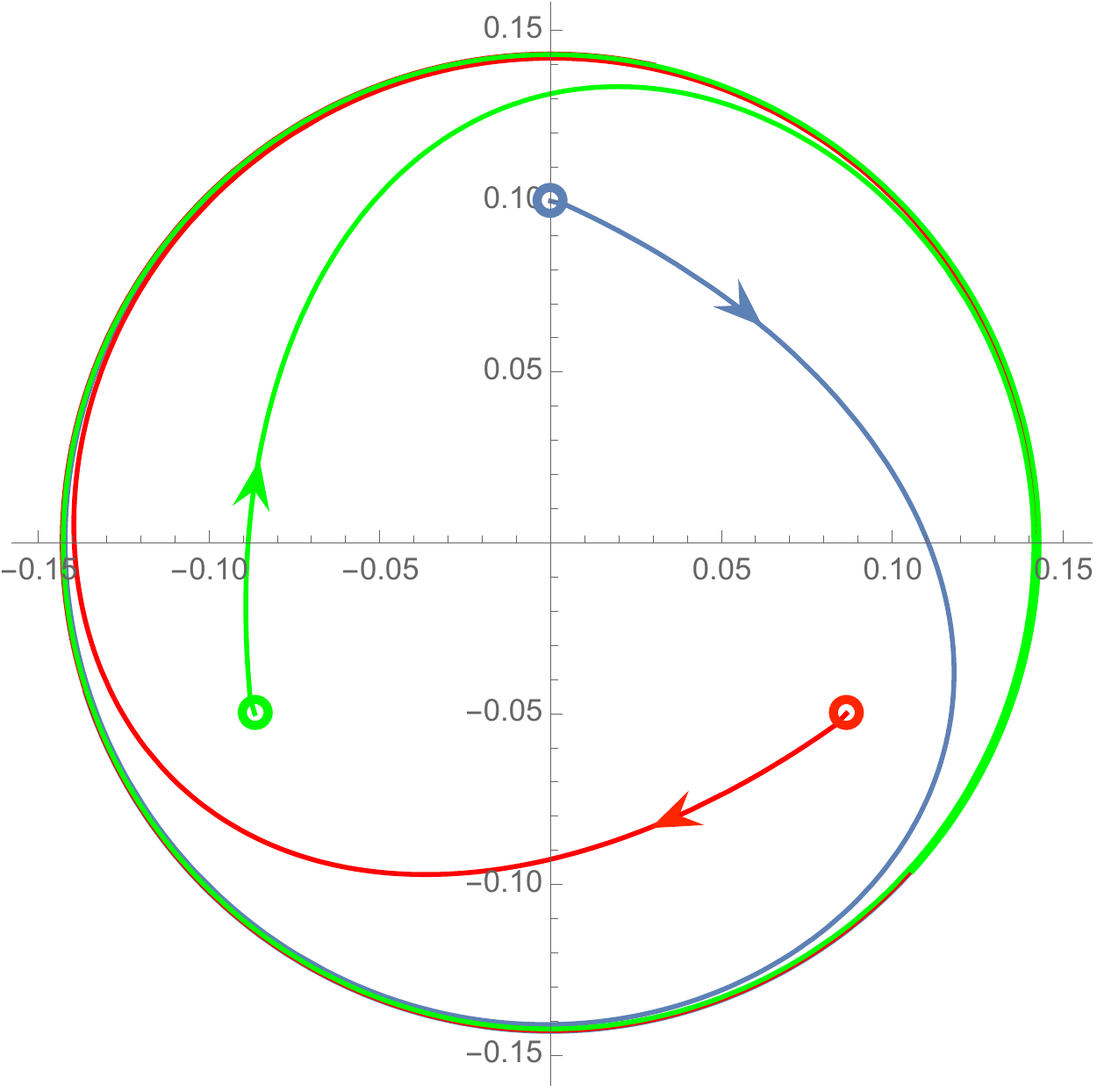}
\caption{\label{fig:spiral}{The solution of the ODE-system \eqref{eq:odes} with the focal particle (blue circle) at initial position $\vec{r}(0)=(0,0.1)$ with $\vec{v}(0)=(0,0.35)$. The particles spiral out towards the equilibrium radius $R_0$.}}
\end{figure}

We solve this system numerically for the the case $N=3$ (see fig.\ \ref{fig:spiral}), and observe two distinct phases: (i) an accelerating phase, where the particles move onto a circle with radius slightly smaller than the equilibrium distance of the potential, and (ii) a milling phase in which the particles in unison slowly spiral outwards towards the equilibrium circle with radius $R_0$. %However, the motion towards the equilibrium position is slow compared to duration of the acceleration phase. 

From the equation system \eqref{eq:odes} and figure \ref{fig:schema}, we can obtain an analytic expression of how the speed decays in the milling phase in the following way.
The second equation in \eqref{eq:odes} gives that
\[\dot{v}_\theta = - F_\theta\]
where $v_\theta$ is the tangential velocity.
By the similarity of the two right angled triangles in figure \ref{fig:schema} we have that
\[F_\theta = \frac{F(R(t))}{R(t)} v_\theta \tau.\]
Note that both $F(R(t))$ and $v_\theta$ tend to zero as $t \rightarrow \infty$. We are interested in studying how quickly that happens, and therefore assume that the system has reached the milling phase. This means that the particles are orbiting very close to the equilibrium circle and that $F(R(t))\approx F_r$ and that $R(t) \approx R_0$. By making these assumptions, we can simplify the differential equation in the following way:
\[\dot{v}_\theta = - F_\theta = - \frac{F(R(t))}{R(t)} v_\theta \tau \approx - \frac{F_r}{R_0} v_\theta \tau = - \frac{v_\theta^2}{R_0^2} v_\theta \tau,\]
where the last equality holds because the particles are following a circular movement with radius $R_0$ for which we have $F_r = v_\theta^2/R_0$. Thus, we end up with the following asymptotic separable differential equation
\[ \dot{v}_\theta = - \frac{v_\theta^3 \tau}{R_0^2} \]
with solution
\begin{equation}\label{eq:odesol}
v_\theta(t)= \frac{1}{\sqrt{\frac{2 \tau t}{R_0^2} + \frac{1}{v_\theta(0)^2}}}.
\end{equation}

\begin{figure}[!htb]
\centering
\includegraphics[width=\columnwidth]{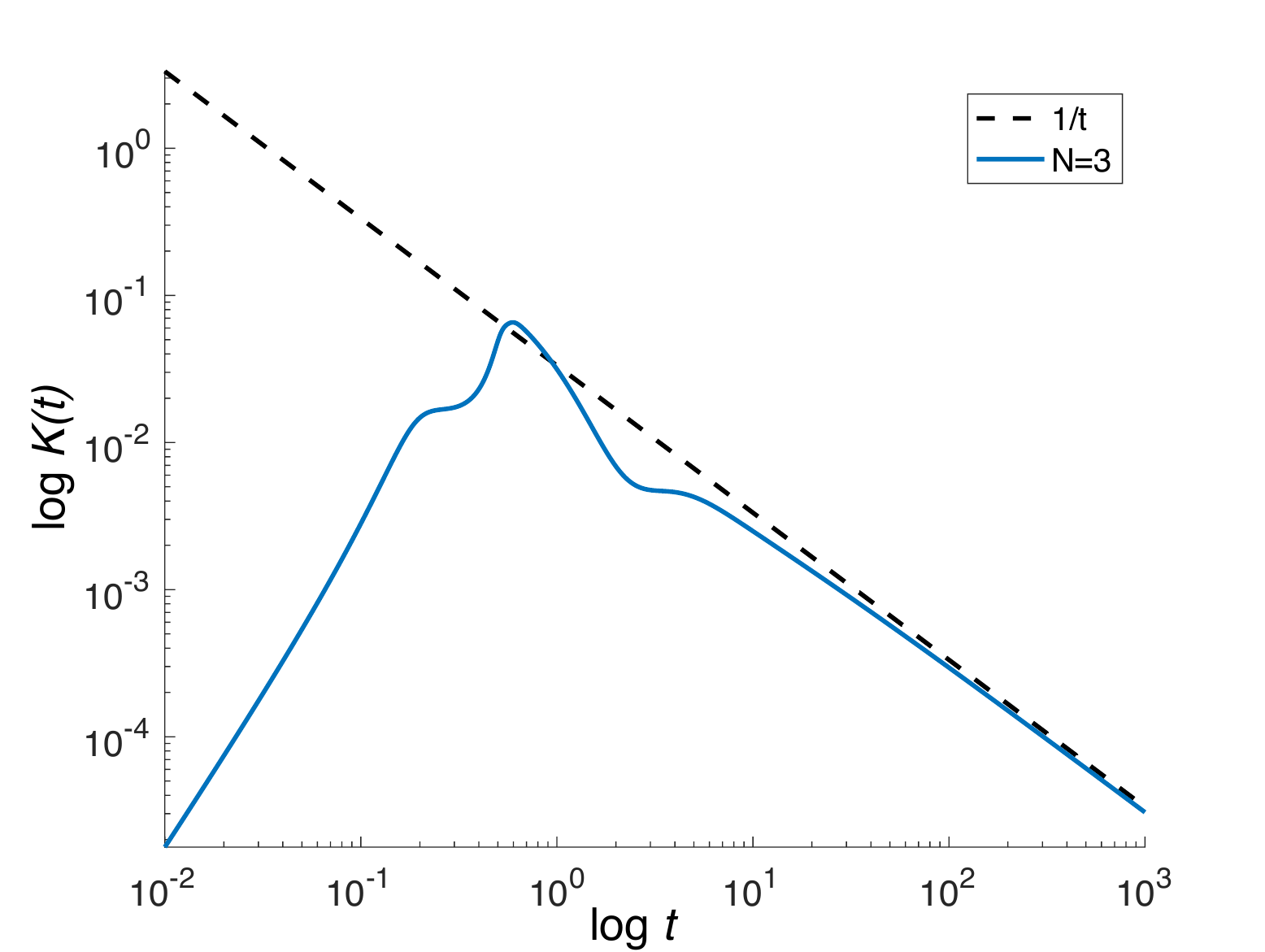}
\caption{\label{fig:kinetic3}{The kinetic energy $K(t)$ as a function of time for $N=3$. The dashed line corresponds to $1/t$. After an initial acceleration the kinetic energy decays approximately as $1/t$ in agreement with \eqref{eq:odesol}.}}
\end{figure}

From this expression it is clear that $|v| \sim t^{-1/2}$ as $t \rightarrow \infty$, and from that we conclude that the kinetic energy should go as $K(t) = m|v|^2/2  \sim t^{-1}$. This result is in agreement with the long-term kinetic energy calculated for a three particle system with random initial conditions (see fig.\ \ref{fig:kinetic3}). This indeed shows that the milling phase is transient, but also that the rate of energy dissipation is low and the kinetic energy scales as $1/t$.

% \begin{figure}[!htb]
% \centering
% \includegraphics[width=\columnwidth]{./VR.pdf}
% \caption{\label{fig:VR}{The relationship between particle velocity $v=|\vec{v}|$ and radius of rotation for $N=3$ obtained with $\tau=1$ after $t=10$ time steps for 100 independent simulations. The rotation typically slows down and the system follows the $(v,R)$-relationship given in \eqref{eq:VR} until it eventually halts at $R=a/\sqrt{3} \approx 0.1443$, where $a=0.25$ is the equilibrium distance of the potential }}
% \end{figure} 

%particles slow down over time we conclude that the radius decreases and that in the limit $t\rightarrow \infty$ all particles will be stationary and at a distance $a$ from one another. 

%For $N=5,7,8$ we also observe the formation of milling structures, but in these cases the structure is asymmetric. This can be understood by considering the stationary configuration in the limit $t\rightarrow \infty$. These limiting configurations with $v_i=0$ for all particles, must necessarily be stationary states of the classical dynamical system with $\tau=0$. It can be shown that in the classical system no circular configurations of particles are stable except for $N=3,4,6$ (\textbf{Visa detta!}). This implies that milling structures for all other $N$ must be either asymmetric or involve more than one circle. For $N=5,8$ we observe a asymmetric rotating configurations, whereas for $N=7$ we see six particles rotating around a stationary particle. 

\subsection{Large particle numbers}
\begin{figure}[!htb]
  \centering
  \includegraphics[width=\columnwidth]{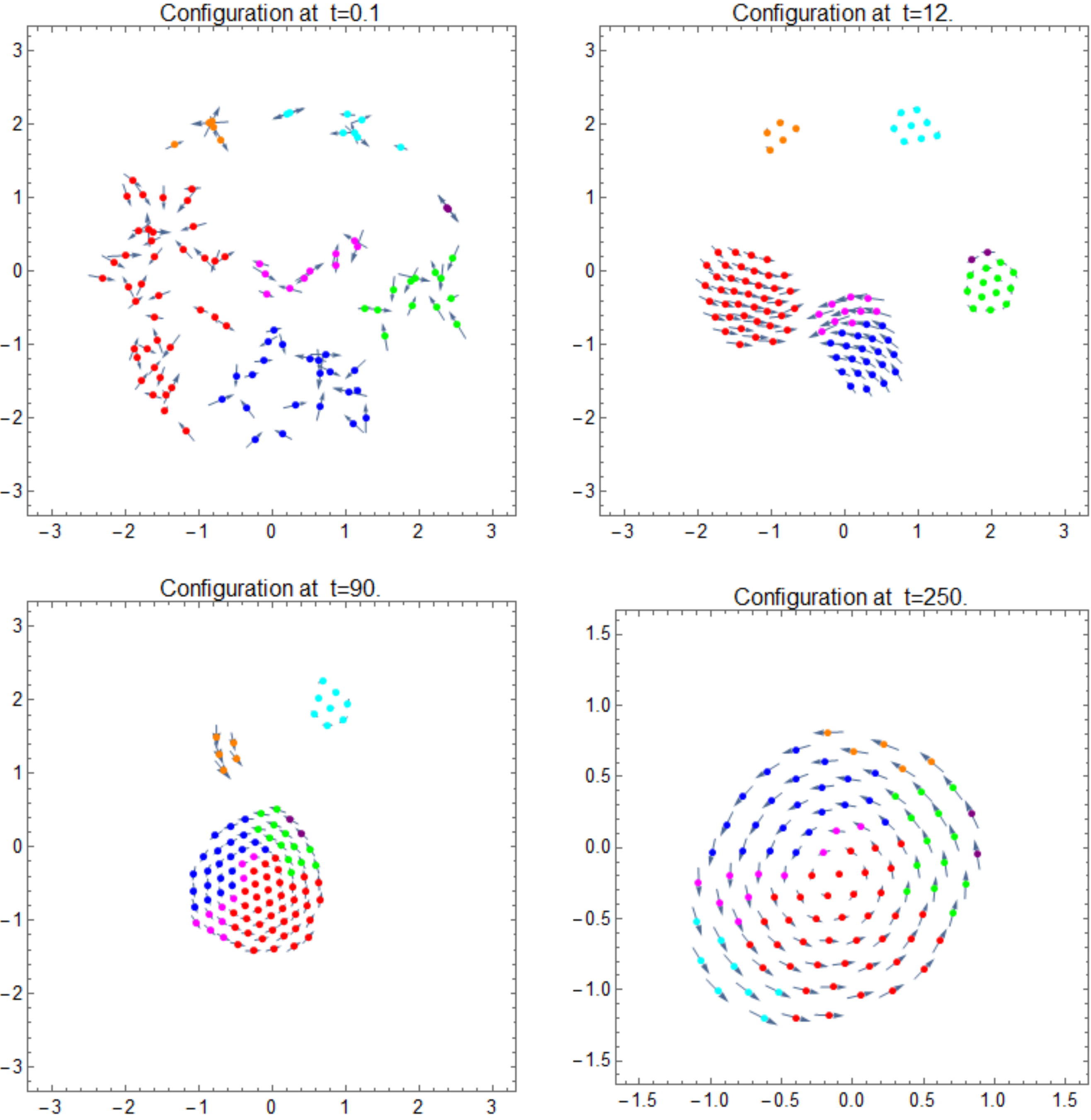}
  \caption{Positions of 100 particles at different times, starting from a
  random initial configuration with zero velocities. The initial
  distribution was uniform in a disk with radius 2.4. The anticipation time $\tau=0.25$, and
the transient time scale equals 1. The particles are coloured by the cluster
they belong to at $t=12$.}
  \label{fig:1_4a}
\end{figure}

\begin{figure}[!htb]
  \centering
 \includegraphics[width=\columnwidth]{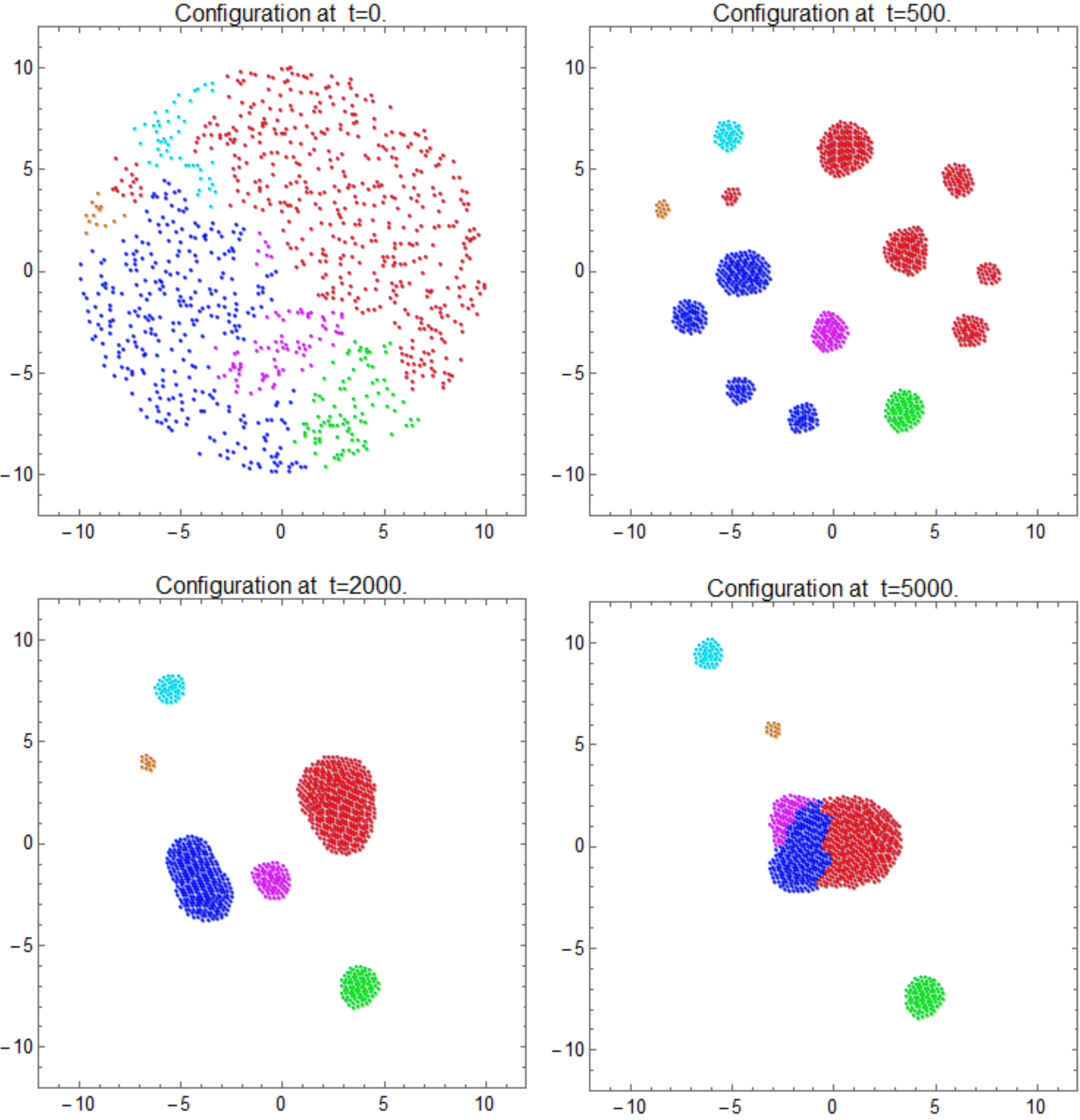}
  \caption{Positions of 1000 particles different times, starting from a
  random initial configuration with zero velocities. The initial
  distribution was uniform in a disk with radius 10.0. The anticipation time $\tau=0.25$, and
the transient time scale equals 0.8. The particles are coloured by the cluster
they belong to at $t=2000$.}
  \label{fig:1_3a}
\end{figure}

We now consider systems with $N\gg10$ particles and look for large scale patterns in the dynamics. Figure \ref{fig:1_4a} contains snapshots from a simulation with $N=100$ particles which show that the particles first aggregate into local clusters ($t \approx 12$) that eventually coalesce and form a single milling structure ($t \approx 90$). Similar dynamics are seen with $N=1000$ particles initialised at the same density (see fig. \ref{fig:1_3a}), with the difference that not all clusters merge within the duration of the simulation. Although not visible due to the large system size, the particles are organised in a hexagonal lattice.

\begin{figure}[!htb]
  \centering
 \includegraphics[width=\columnwidth]{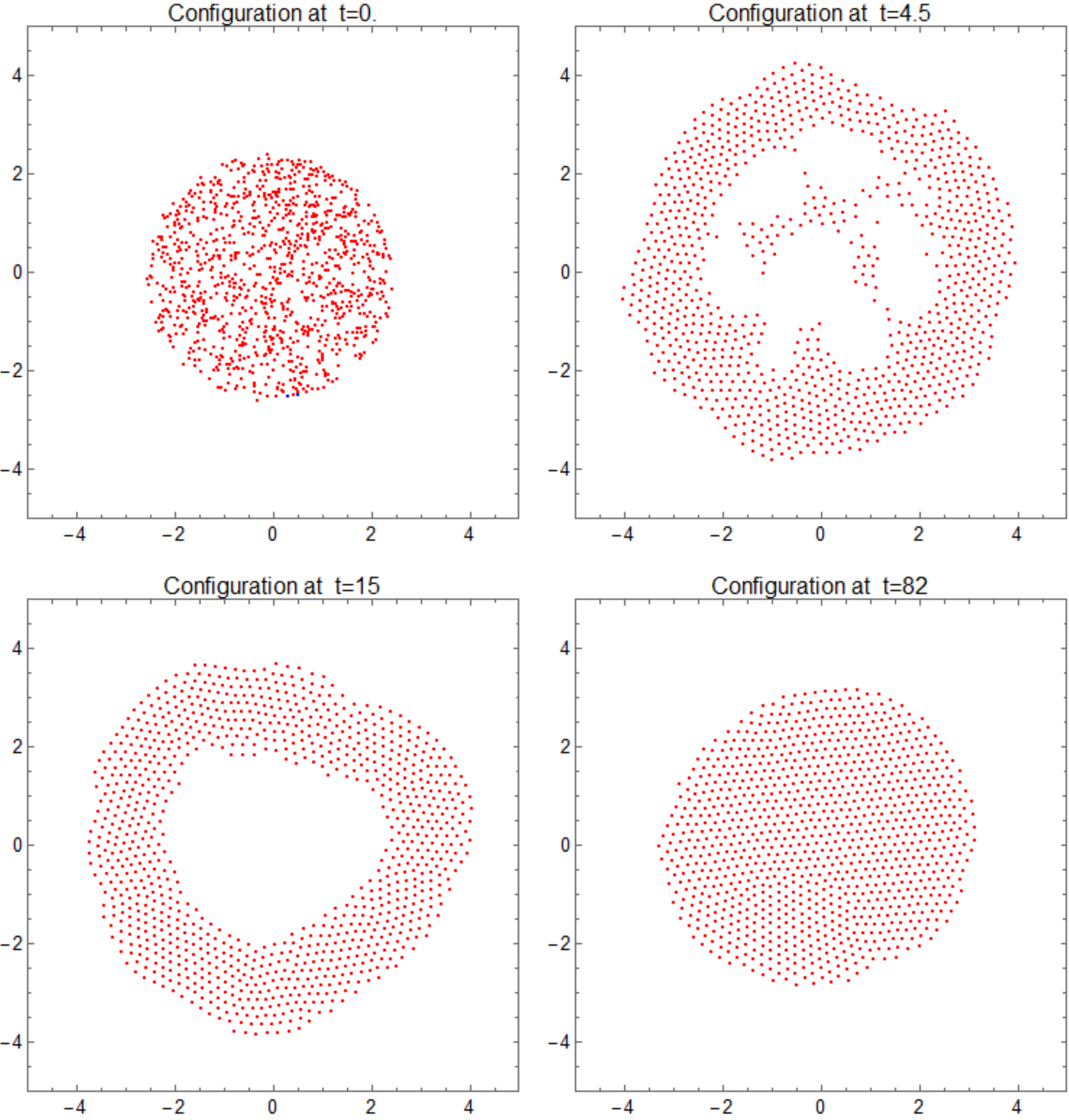}
  \caption{Positions of 1000 particles different times, starting from a
  random initial configuration with zero velocities. The initial
  distribution was uniform in a disk with radius 2.5. The anticipation time $\tau=0.05$, and
the transient time scale equals 0.06. }
  \label{fig:1_2a}
\end{figure}

In the case of $N=1000$ particles initialised at a $10$-fold higher density (see fig.\ \ref{fig:1_2a}) we observe an initial repulsive phase (at $t \approx 4.5$), which leads to an expansion of the cluster (at $t \approx 15$). However, the cluster remains coherent and contracts into an approximately hexagonal lattice, which rotates around the centre of gravity. In conclusion, we observe similar dynamics independent of system size. The system which is initially disordered self-organises into an approximately hexagonal lattice which rotates as a rigid body around its centre of gravity. 

\begin{figure}[htb!]
  \centering
\includegraphics[width=\columnwidth]{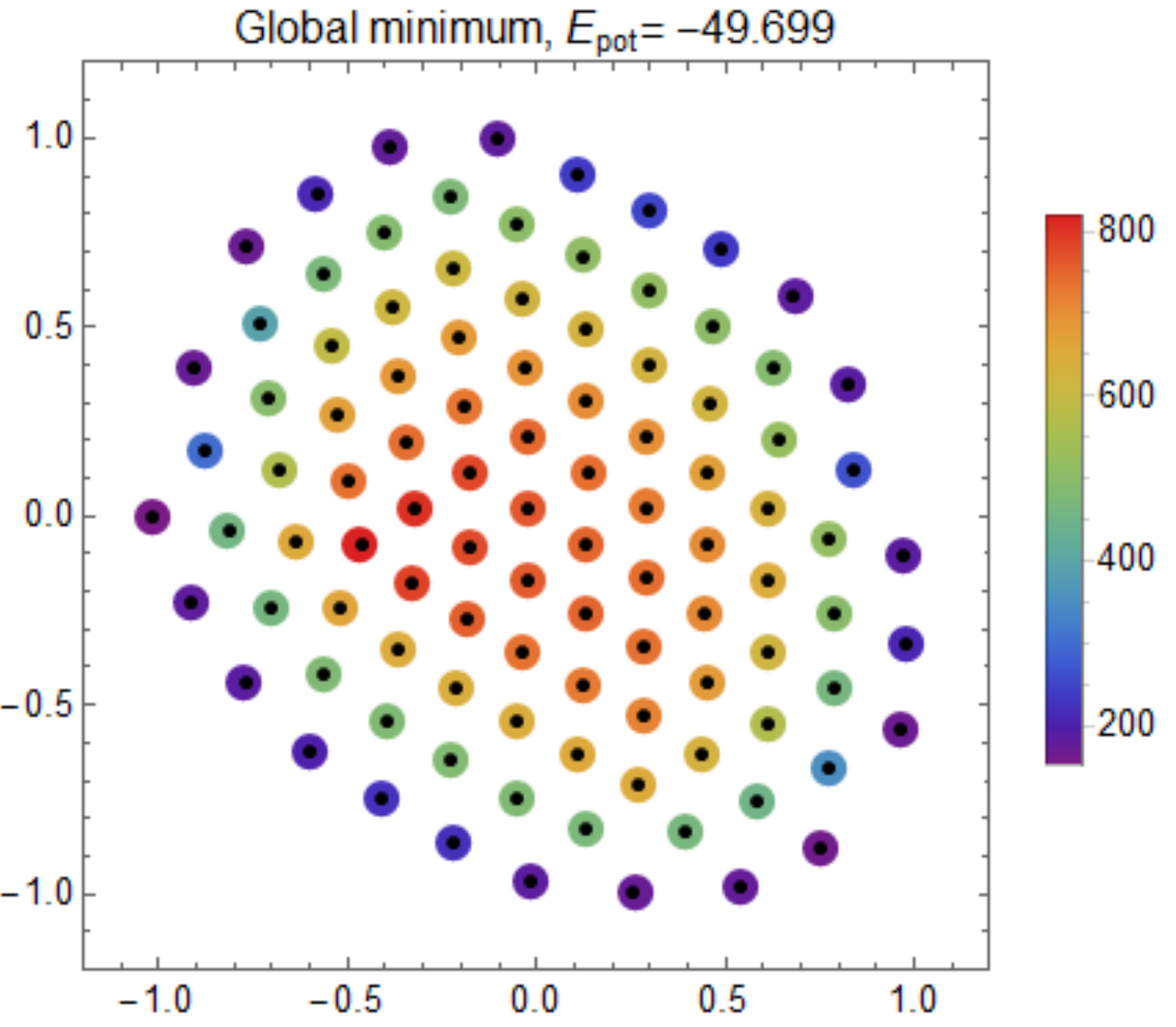}
  \caption{The black dots show the final point configuration, and the coloured dots  a local energy minimum. The colours represent the stiffness of each particle position, calculated from the largest eigenvalue of the linearised system as a function of only one particle, given that the others are fixed.}
  \label{fig:1_1d}
\end{figure}

It appears as if the inclusion of anticipation allows the system to dissipate energy and reach a state which minimises the total potential energy. 
In order to investigate this we compared the configuration in figure \ref{fig:1_4a} at $t=250$ with a configuration of stationary particles that minimises the total potential energy (calculated using Mathematica's function \texttt{Minimize}). The comparison is shown in figure \ref{fig:1_1d} and reveals almost perfect agreement between the milling configuration obtained with anticipation (black dots) and the stationary configuration that minimises the total potential energy, which consists of a partial hexagonal lattice. 

We now move on to the question of energy dissipation for large $N$, which seems to play a crucial role in the dynamics of the system.

%The colours correspond to the stiffness of each particle position, calculated as the largest eigenvalue of the linearised system.

%Simulations in 3 dimensions exhibit a similar behaviour, where for intermediate $\tau$ the particles aggregate and move on concentric spheres (data not shown).

\subsection{Energy dissipation}\label{sec:diss}
{\color{black} 
The example with two spring-coupled particles, and the special case of a symmetric configuration of three particles  shows that anticipation may result in energy dissipation. The following calculation shows that the total energy, $E(t)$, computed as the sum of kinetic energy and the anticipated potential energy,
\begin{align}
E(t) = \sum_{i=1}^N |\vec{v}_i|^2 + \sum_{\substack{i,j=1\\i\ne j}}^N U\left( \vec{x}_i-\vec{x}_j +\tau (\vec{v}_i-\vec{v}_j)\right)\,,
\end{align}
is strictly decreasing along the orbits, when not all velocities are constant. 
Here we consider a more general anticipating dynamical system, of which the
particle system studied above is but a special case: 
\begin{align}
  \label{eq:antsys}
  \dot{\vec{x}}&= \vec{v} \nonumber\\
  \dot{\vec{v}} &= -\nabla U( \vec{x} + \tau \vec{v})\,,
\end{align}
where $\vec{x}, \vec{v} \in \R^n$, and $U: \R^n\rightarrow \R$ is a
smooth and bounded potential.  Then
\begin{align}
 \label{eq:antp}
  \frac{d}{dt}\left( E(t) \right) &= \frac{d}{dt}\left( \frac{|\vec{v}|^2}{2} +
     U(\vec{x}+\tau\vec{v})\right) \\
     &=  \vec{v}\cdot\dot{\vec{v}} +
                      \nabla U(\vec{x}+\tau\vec{v}) \cdot
                      \left(\dot{\vec{x}} +\tau \dot{\vec{v}}\right)
                                     \nonumber\\
  &= \vec{v} \cdot \underbrace{\left( \dot{\vec{v}} +
    \nabla U(\vec{x}+\tau\vec{v})\right)}_{=0} -\tau | \dot{\vec{v}}|^2\,.
\end{align}
Hence the sum of kinetic energy and the anticipated potential energy
are strictly decreasing until all velocities are constant. This behaviour can be seen in figure \ref{fig:kinetic}A, which shows the how the total energy evolves for the simulation presented in figure \ref{fig:1_4a} with $N=100$ particles. The anticipated total energy is strictly decreasing and the jumps correspond to collision and merger of clusters. Note that the classical total energy (calculated with $\tau = 0$) is not strictly decreasing but increases transiently during collisions. 

\begin{figure}[hbt!]
  \centering
 \includegraphics[width=\columnwidth]{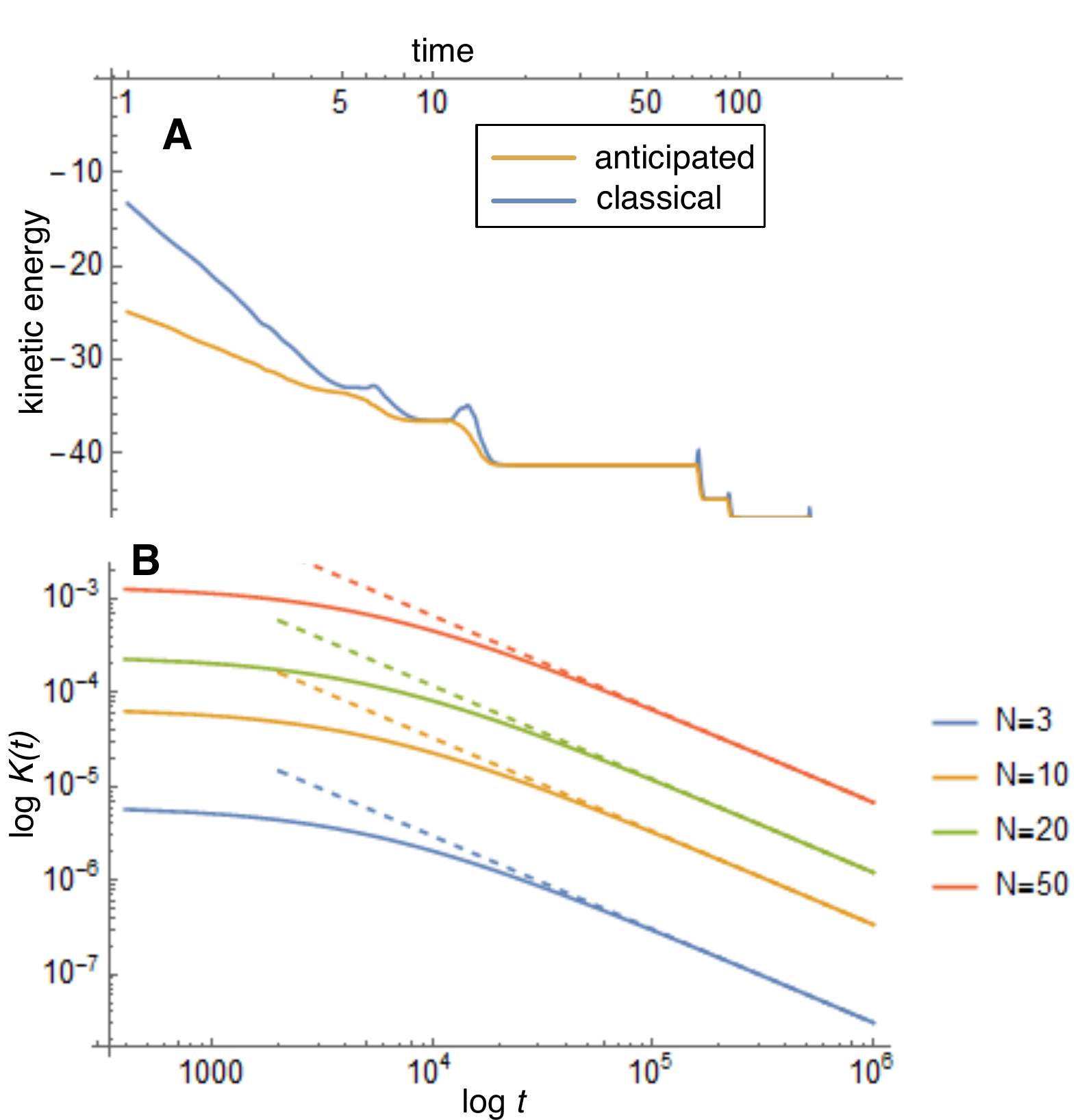}
 \caption{(A) The evolution of energy for the simulation represented in fig~\ref{fig:1_4a}. The jumps in
  the two curves correspond to the merger of clusters. (B) The evolution of the kinetic energy
  for $N=3,10,20$ and $50$. The
  initial condition was chosen with positions according to a global minimum of
  the potential, and velocities corresponding to a rigid rotation of
  the point configuration. The slope of the dashed lines is $-1$.} 
  \label{fig:kinetic}
\end{figure}

\subsubsection{The rate of dissipation}\label{sec:dissrate}
For the three particle system we could show that the energy decays as $1/t$. We now analyse the rate of energy dissipation for a general potential $U(x)$ and arbitrary system size $N$. 

With a potential given by a positive definite quadratic form, $U(x) =
\vec{x}^T A \vec{x}$, the system is again a damped oscillator, whose
solutions converge exponentially to zero, but for a semidefinit form,
say
$U=\frac{1}{2}\lambda_2 x_2^2+\cdots + \frac{1}{2}\lambda_n x_n^2$,
the system never comes to rest, because the potential is constant
along the degenerate $x_1$-direction. The anticipation does not change
this, and seen 
from the point of view of simulations, the same will be observed if
the quadratic form has an almost degenerate subspace, in the sense
that some of the eigenvalues of $A$ are several magnitudes smaller than
the the dominant eigenvalues.

However, if the potential $U$ is constant, or nearly constant, on a
curved manifold, e.g. a curve, the situation is different.
The three particles on a circle is an example of this more general
situation. Here the potential attains
its minimum on a one-dimensional curved submanifold, the circle, with
 $\vec{x}(t)$ very close to the circle, and with $\vec{v}(t)$
almost tangential to the manifold. A Taylor expansion of the potential
$U$ then gives
\begin{align*}
  \frac{d}{dt}\vec{v}_{\parallel}(t) = \bigoh(|\vec{v}_{\parallel}|^2)
\end{align*}
at least if $\tau |\vec{v}_{\parallel}|$ is small compared to the
curvature of the manifold. This calculation can be carried out in a
more general 
setting, for example when the minimum of $U$ is attained along a curve
 in $\R^n$: Assume
that $\Gamma=\{ \vec{\phi}(x)=(\phi_1(s),...,\phi_n(s))\,|\, s\in
  I\subset\R\}$ is a 
segment of a curve parametrised by $s$, and assume that $U(x)$ is a
quadratic function of $\vec{x}- \vec{x}^*$, where $\vec{x}^*$ is the point on
$\Gamma$ closest to $x$. Though not completely general, this would
cover many generic situations. We may choose a coordinate system so
that
\begin{align*}
  \phi_k(0) &= 0\;\mbox{for} \; k=1,...,n\,,\\
  \phi_1(s) &= s,\\
  \phi_k'(0) &= 0\;\mbox{for} \; k=2,...,n\,,\\
  \phi_k''(0) &= \mu_k\;\mbox{for} \; k=1,...,n\,,
\end{align*}
so that the curve is tangent to the $x_1$-axis of the coordinate
system, and that it may be parameterised with $x_1$. The point
$\vec{x}^*=(s, \phi_2(s),...,\phi_n(s))$ is found by
minimising the function $\sum (x_k-\phi_k(s))^2$, and at a minimum,
this must satisfy
\begin{align}
  \label{eq:smin}
  s-x_1+\sum_{k=2}^n ( \phi_k(s)-x_k)\phi_k'(s)=0\,,
\end{align}
which determines $s$ as a function of $\vec{x}$, at least near the
origin. Hence
\begin{align*}
  U(x_1,...,x_n)&= \frac{1}{2}(\vec{x}-\vec{\phi}(s))^T A
                  (\vec{x}-\vec{\phi}(s)) \\
  &= \frac{1}{2}\nu_1 (x_1-s)^2 +\frac{1}{2}\sum_{k=2}^n \nu_k
    (\phi_k(s)-x_k)^2\,. 
\end{align*}
where for simplicity we have assumed that $A$ is diagonal with diagonal
elements $\nu_k$. Restricting Equation~(\ref{eq:smin}) to points on the $x_1$-axis, we
find that
\begin{align*}
  x_1(s) &= s +\sum_{k=2}^n \frac{1}{2}\mu_k s^3
           +\bigoh(s^4)\,\qquad\mbox{or}
  \\
  s(x_1) &= x_1 -\sum_{k=2}^n \frac{1}{2}\mu_k x_1 ^3
           +\bigoh(x_1^4)\,,
\end{align*}
so that
\begin{align*}
  U(x_1,0,...,0)&= \bigoh(s^6)+ \frac{1}{2}\sum_{k=2}^n \nu_k
                  \left( \frac{1}{2}\mu_k s^2 +\bigoh(s^3)\right)^2\\
  &= \frac{1}{2}\sum_{k=2}^n\sum_{j=2}^n\mu_j\mu_k\nu_k x_1^4+\bigoh(x_1^5)
\end{align*}
Therefore, with $\vec{v}=(v_1,0,...,0)$ and $\vec{x}=\vec{0}$,
\begin{align*}
  \frac{d}{dt}\vec{v} &= -\nabla U(\vec{x}+\tau \vec{v}) \;
                       = -\nabla U(\tau v_1,0,...,0)\,,  
\end{align*}
and finally
\begin{align}
  \label{eq:oneovert}
  \frac{d}{dt}\frac{|\vec{v}|^2}{2} &= - \vec{v}\cdot \nabla U(\tau v_1,0,...,0) \\
  &= -\gamma v_1 \tau^3 v_1^3 + \bigoh(v_1^5) \\
  &= -\gamma \tau^3 v_1^4 + \bigoh(v_1^5) \,,
\end{align}
where $\gamma=2\sum_{k=2}^n\sum_{j=2}^n\mu_j\mu_k\nu_k$,
and therefore ${\frac{d}{dt}|\vec{v}|^2 \sim - |\vec{v}|^4}$, which
implies that
\begin{align*}
   |\vec{v}(t)|^2 &\sim \frac{1}{t}
\end{align*}
when $t$ is large, and the exact expression depends on the
constants $\mu_k$, which determine the curvature of the curve.
Of course this is not a full proof, but rather motivation for the
observed behaviour. A complete proof would require an additional
calculation showing that the the component of $\vec{F}$ normal to
$\Gamma$ is sufficiently strong to confine $\vec{x}(t)$ to remain
sufficiently close to $\Gamma$, and $\vec{v}(t)$ to rest almost
tangent to $\Gamma$, that the estimates above are still valid. 
%Therefore
%\begin{align*}
%  \frac{d}{dt}\frac{|\vec{v}|^2}{2} &= \vec{v}\cdot \vec{F}(\tau
%                                      \vec{f}) \\
%  &= -\gamma v_1 \tau^3 v_1^3 + \bigoh(v_1^5) \,,
%\end{align*}
%where $\gamma=\sum_{k=2}^n\sum_{j=2}^n\mu_j\mu_k\nu_k$,
%and therefore $\frac{d}{dt}|\vec{v}|^2 \sim - |\vec{v}|^4$, which
%implies that
%\begin{align*}
%   |\vec{v}(t)|^2 &\sim \frac{1}{t}
%\end{align*}
%when $t$ is large, and the exact expression depends on the
%constants $\mu_k$, which determine the curvature of the curve. If the
%particle is not exactly on $\Gamma$, but at   
%a small distance, a similar calculation would reveal a force bringing
%$\vec{x}$ rapidly towards $\Gamma$.
%
%%It is possible to express this in more geometrical terms, which may be
%useful for understanding situations with a minimizing manifold of
%higher dimension, but it seems diffiult to make a very concise proof of
%the statement in that way.

For the $N$-particle systems with  
pair interactions that is the main theme of this paper, the potential
energy is invariant under rotations and translations, and hence the
potential miniumum is attained on a three-dimensional curved
submanifold of the $2 n$-dimensional configuration space. In
simulations with a large number of particles it is difficult to
observe the $1/t$ behaviour, however, because the dynamics is complex
and one may need to wait a very long time before this asymptotic
behaviour can be seen. To circumvent this problem we have computed numerical solutions with
initial data such that the positions $\vec{x}_i(0)$ correspond to a
minimum for the potential energy, and with initial velocities corresponding to
a rigidly rotating body, and find excellent agreement with the
predicted $1/t$-decay of energy (see fig. \ref{fig:kinetic}B).

Situations where the mimimum manifold has higher dimension appear in
simulations with a large number of particles, where we often observe
how the system breaks up into smaller clusters that 
essentially do not interact with the other clusters, and hence
represent a case with a many dimensional subspace with almost 
constant potential energy, defined by translations and rotations of
each cluster.

%\textbf{Spara detta?}
%Note that in the above calculation we used the anticipated positions $\vec{x}^p$ when calculating the potential energy (since it is the anticipated positions that enter into the interparticle forces). If the potential energy instead is calculated from the current position we observe that the total energy may transiently increase when clusters merge (see fig. \ref{fig:energy}). This observation suggests that the anticipated potential energy is preferable when characterising the dynamics of the system, since no external energy is added to the system. 

\subsection{Linear and angular momenta}
So far we have shown that anticipation induces dissipation and generally leads to a $1/t$-decay in kinetic energy. This leads to a gradual reduction in the rotation of the cluster, but we have not explained how this rotation comes about in the first place. To do this we need to analyse the time evolution of the linear and angular momenta, $\vec{M}(t)$ and $L(t)$, which we define as
\begin{align}
  \vec{M}(t) &= \sum_{i=1}^N \vec{v}_i  \qquad\quad \mbox{and} \qquad L(t) =
         \sum_{i=1}^N \vec{v}_i \times \vec{x}_i\,,
\end{align}
where $\vec{v}_i \times \vec{x}_i$ denotes the cross product of
$\vec{v}$ and $\vec{x}$, which is a scalar in this two
dimensional setting. First,
\begin{align}
  \frac{d}{dt}\vec{M}(t)&= \sum_{i=1}^N \frac{d}{dt} \vec{v}_i \\
                  &= \sum_{\substack{i,j\\i\ne j}} F_{i,j} =
  \frac{1}{2} \sum_{\substack{i,j\\i\ne j}}
  \left(F_{i,j}+F_{j,i}\right) =0\,.    \nonumber
\end{align}
By a change of variables we may assume that $M(t)=\vec{0}$, and hence
that the center of mass, $\frac{1}{N}\sum_{i} \vec{x}_i$ is constant,
and may therefore be set equal to zero.  The angular momentum
satisfies
\begin{align}
  \frac{d}{dt}L(t) &= \sum_i \left(\dot{\vec{v}}_i\times \vec{x}_i + 
                     \vec{v}_i\times \dot{\vec{x}}_i \right)  \\
                   &= \sum_{\substack{i,j\\i\ne j}} F_{i,j} \times
 \vec{x}_i = \frac{1}{2}\sum_{\substack{i,j\\i\ne j}} F_{i,j} \times
  \left(  \vec{x}_i-\vec{x}_j\right). \nonumber
\end{align}
Note that the force $F_{i,j}$ may be written
\begin{align}
  F_{i,j} &= - \frac{U'\left(\left|\rrj\right|\right)}
            {\left|\rrj \right|}\rrj \equiv
           - V(\left|\rrj \right|) \rrj ,
\end{align}
where $\rrj=\vec{x}_i-\vec{x}_j+\tau(\vec{v}_i-\vec{v}_j)$, and we define $V(r) = U'(r)/r$.
%and where $U(r)$ is the real function that defines the Morse potential.
Therefore
\begin{align}
 \frac{d}{dt}L(t) &= - \frac{1}{2}\tau \sum_{\substack{i,j\\i\ne
  j}}V(\left|\rrj\right|)(\vec{v}_i-\vec{v}_j)
  \times
  \left(  \vec{x}_i-\vec{x}_j\right)
\end{align}
So with $\tau=0$, the angular momentum is constant as it should for a classical system. All
simulations presented here start with all initial velocities equal to
zero, and therefore also $\dot{L}(0)=0$, but the second derivative
need not be. Computing $\ddot{L}(t)$ gives
\begin{align}
  \frac{d^2}{dt^2}L(t)&= \sum_{\substack{i,j\\i\ne
  j}}\Bigg(V'(\left|\rrj\right|)\frac{\rrj\cdot
  \rrjd}{\left|\rrj\right|}\left(\vec{v}_i-\vec{v}_j\right)\times\left(\vec{x}_i-\vec{x}_j
  \right)    +\\
                      &
                        \qquad\qquad 
    V(\left|\rrj\right|)\left(\dot{\vec{v}}_i-\dot{\vec{v}}_j\right)\times\left(\vec{x}_i-\vec{x}_j
  \right)   \Bigg).  \nonumber
\end{align}
At $t=0$, when the velocities are zero, the first terms in the sum
disappear, and using the symmetries in the sum this yields
\begin{align}
  & \frac{d^2L}{dt^2}(0)=\\
   &= \frac{\tau}{2}\sum_{\substack{i,j,k\\i\ne j\ne k}} V(|\vec{x}_i-\vec{x}_j|)
  V(|\vec{x}_i-\vec{x}_k|)(\vec{x}_i-\vec{x}_j)\times  (\vec{x}_i-\vec{x}_k)
\end{align}
which typically will be non-zero for a random initial
configuration. Therefore, contrary to  conservative systems,
asymmetries in the initial configuration of a particle system
initially at rest, may result in a non-zero angular momentum, which is
then observed as a milling structure. 

The time evolution of the angular momentum of a typical simulation is exhibited in figure \ref{fig:angular}A which shows $L(t)$ for the simulation presented in figure \ref{fig:1_4a}. The jumps in the curve correspond to the merger of clusters, and we note that collisions typically increase the absolute value of the angular momentum, while it decreases in between collisions due to the $1/t$ dissipation of kinetic energy discussed above. Since $L(t)$ is proportional to the speed $v(t)$, which scales as $1/\sqrt{t}$, we expect the angular momentum to asymptotically scale as $1/\sqrt{t}$, which is precisely what is observed in simulations (see fig.\ \ref{fig:angular}B).

\begin{figure}[htb!]
  \centering
\includegraphics[width=\columnwidth]{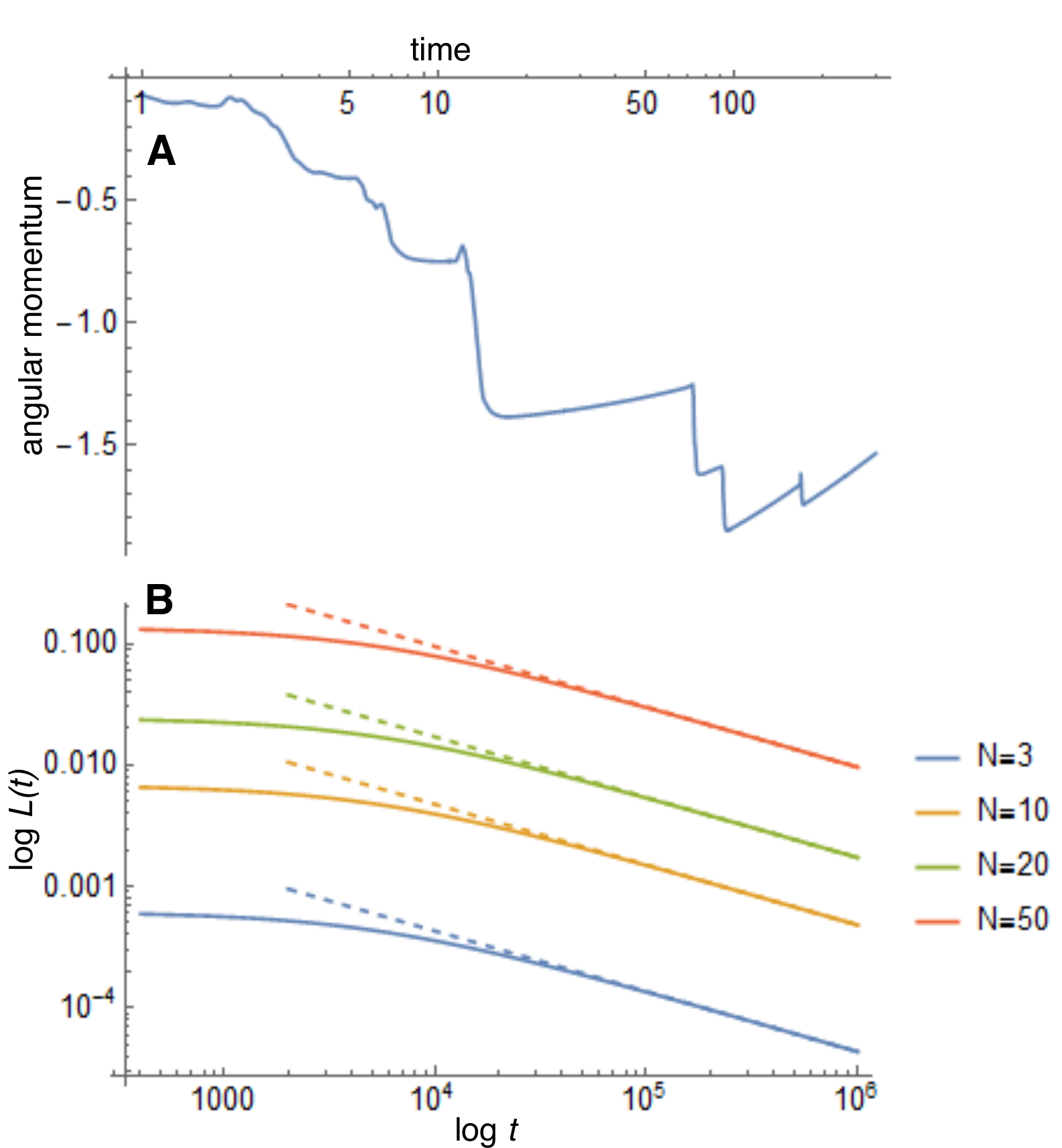}
  \caption{(A) The evolution of angular momentum for the simulation represented in fig~\ref{fig:1_4a}. The jumps in the curve correspond to the merger of clusters. (B) The evolution of the angular momentum for $N=3,10,20$ and $50$. The
  initial condition was chosen with positions according to a global minimum of
  the potential, and velocities corresponding to a rigid rotation of
  the point configuration. The slope of the dashed lines is $-1/2$.} 
  \label{fig:angular}
\end{figure}

\section{Discussion}
In this paper we have shown that internal prediction or anticipation leads to milling behaviour in a system of particles that interact via a pairwise potential. The behaviour of the system depends on the relation between the anticipation time $\tau$ and the transient time scale $\bar{t}_T$ \eqref{eq:tt}, such that when $\tau$ is of the same order of magnitude as $\bar{t}_T$ we observe the rapid formation of rotating clusters that merge upon collision. These clusters have a crystalline structure made up of a hexagonal arrangement of particles and rotate around the centre of gravity as a rigid body. We have shown analytically that rotation emerges due to an increase in the angular momentum, which only occurs for $\tau >0$. Further we have shown that anticipation leads to dissipative behaviour where the kinetic energy typically decreases as $1/t$.

These results show that motion anticipation -- the capability to extrapolate the future location of their peers -- which is known to be present in humans and other animals, has a large impact on the dynamics of interacting particle models that are often used for describing collective behaviour. It acts to stabilise the system, which leads to an ordered crystalline structure, and also induces milling which is a ubiquitous feature of flocking -- even in the absence of self-propulsion. 

Using models that incorporate anticipation it would be possible to analyse existing data on e.g. pedestrian to infer the anticipation time $\tau$ that is used by humans \cite{Karamouzas2014}. This could then be compared with anticipation time obtained from other species and other systems, and could give novel insight into how anticipation influences collective behaviour at different spatial and temporal scales. 

In relation to experimental data we note that the fixed value of $\tau$ most likely is a simplification, and that most animals (including humans) make use of a dynamical value of $\tau$ which depends on the distance to the object whose trajectory one is trying to predict.
For larger distance a larger value of $\tau$ is beneficial, while for smaller distances one runs the risk of over-shooting unless $\tau$ is chosen sufficiently small. In terms of the interacting particle systems we have studied here we conjecture that including a dynamic anticipation time will lead to a faster equilibration of the system dynamics.

A possible extension to the current model is to take into account the fact that an individual might have knowledge about the predictions made by other individuals. That knowledge could be captured in an internalised model that would give individual predictions that go beyond the simple linear extrapolation used in the current model. This would lead to a kind of second order prediction that might lead to other dynamical behaviours. 

It is also possible to consider a model in which individuals have perfect knowledge about the future state of the system. This implies that ${x}(t+\tau) = x(t)+\tau v(t)$, which when inserted into  eq. \eqref{eq:x} (instead of $x^p(t+\tau)$) would give rise to a reversed delay-equation. Although such a system would contradict the principle of causality it could still be interesting from a theoretical point of view.

The observation that anticipation introduces dissipation and makes it possible for the system to reach steady states 
%allows for non-trivial steady states in a system of interacting particles 
might have implications for our understanding of more complicated dynamical systems. Many biological and sociological systems contain an element of anticipation and models that explicitly take this into account might give novel insights into the dynamics of these systems.

%One might for example argue that stability of human societies depends on our ability to forecast the future and act on this information to maximise some utility. This behaviour is made possible by the human ability to construct internal and external representations and models of systems and subsystems whose future behaviour we are interested in. These systems often consist of other people and hence we are in a situation similar to the one investigated in this Letter, where individuals are making predictions about future states of the system and acting on this information.  

%\bibliography{biblio}

\end{document}